\documentstyle[prl,floats,aps]{revtex}

\begin{document}
\draft

\twocolumn
\begin{title}
{C-axis negative magnetoresistance and upper critical field
of Bi$_{2}$Sr$_{2}$CaCu$_{2}$O$_{8}$. } 
\end{title}
\author{ V.N. Zavaritsky $^{1,2}$, M. Springford $^{1}$, and A.S.
Alexandrov$^{3}$}
\address
{$^{1}$ H.H. Wills Physics Laboratory, University of Bristol, Bristol
BS8
1TL, United Kingdom; $^{2}$ Kapitza Institute for Physical Problems, 2
Kosygina
Str., 117334 Moscow (Russia);
$^{3}$ Department of Physics, Loughborough University, Loughborough LE11
3TU, United Kingdom}
\maketitle
\begin{abstract}
The out-of-plane
resistance and the resistive upper critical field of BSCCO-2212 single crystals with T$_{c0}\simeq 91-93\,K$
have
been measured  in  magnetic fields
up to 50 T over a wide temperature range. The results are 
characterised by a positive linear magnetoresistance in the 
superconducting state and a negative linear magnetoresistance in the 
normal state. The zero field 
normal state c-axis resistance, the negative linear normal
state magnetoresistance,  and the
divergent upper critical field H$_{c2}$(T) are explained
in the framework of the bipolaron theory
of superconductivity.
\end{abstract}
\narrowtext


High magnetic fields have been widely used to explore the single 
particle spectrum of normal and superconducting metals \cite{spr}.  
Historically, de Haas-van Alphen effect oscillations have provided 
precise and detailed information on the Fermi surface and the damping 
of quasiparticles in Landau Fermi liquids.  Such oscillations have 
also been studied in the vortex state of many low-T$_{c}$ type-II 
superconductors \cite{review} yielding information on the electronic 
many-body environment in the non-Fermi liquid BCS state.  In the 
cuprates superconductors, high magnetic field studies have revealed a 
non-Fermi liquid temperature dependence of both ab- and c-axis 
resistivities \cite{and} and a non-BCS divergent shape of the upper 
critical field $H_{c2}(T)$ \cite{mac,oso,alezav,fra,boe,gan}.  These 
studies were performed both in relatively low-T$_{c}$ cuprates 
\cite{mac,oso,fra,boe,gan} and in some high-T$_{c}$ compounds in a 
moderate \cite{alezav} (below 15 T) and high (up to 60 T) field 
(see Ref.\cite{and,boe} and   more recent results  Ref.\cite{brus}).  The upper critical field was determined from the 
temperature dependence of c-axis  resistivity with some uncertainty due to 
fluctuations \cite{alezav}. The uncertainty was removed in the
comprehensive study by Gantmakher $et$ $al$ \cite{gan} of the
in-plane resistivity of high-quality YBa$_2$Cu$_3$O$_{7-\delta}$
crystals, which confirmed the non-BCS upper critical field observed in
Ref. \cite{mac,oso,alezav,fra,boe} strongly  supporting  the
bipolaron theory of cuprates \cite{ale,alemot}.
 
  We report here on a study in pulsed magnetic fields of 
  Bi$_{2}$Sr$_{2}$CaCu$_{2}$O$_{8}$, with T$_{c0}\simeq 91-93\,K$, 
  which reveals new features of the c-axis transport.  We observe a 
  positive linear magnetoresistance in the flux flow 
  (superconducting) regime and a negative linear magnetoresistance in 
  the normal state.  This allows for  determination of the 
  upper critical field,  as the point of intersection of these two 
  regimes.  We have measured $H_{c2}(T)$ 
   as a function of temperature over a 
  wide temperature range, $0.2\leq T/T_{c0}\leq 1$, and find a 
  divergent behaviour consistent with results in other materials 
  \cite{mac,oso,alezav,fra,boe,gan}.  We discuss this from the point of 
  view of the bipolaron theory of superconductivity \cite{ale}.

 BSCCO-2212 single crystals were grown by solid state reaction
 \cite{my-irr}.
 Five samples with 
 in-plane dimensions from $\simeq 85 \times 110 \mu m^2$ to $\simeq 26 
 \times 30 \mu m^2$ and thicknesses of $\approx 1.5-4.3\mu m$ have 
 been thoroughly studied in pulsed fields. All measurements were of
 the longitudinal magnetoresistance with field and current directed
 along the c-axis.  The absence 
 of hysteresis between the data obtained on the rising and falling 
 sides of the pulse, characterised by very different values $\partial 
 B/\partial t$, excludes any significant heating of our samples.  
 
Fig.1 shows a typical measurement of the effect of magnetic field on 
 the out-of-plane resistance of a BSCCO-2212 single crystal below
 a zero-field critical temperature,
 $T_{c0}$.  There is a low-field   regime, $R_{FF}(B,T)$, where a 
 linear field dependence fits the experimental observations rather 
 well, Fig.1. As has been  suggested in Ref. \cite{kos} the origin of  the finite
 c-axis resistivity below the zero field transition 
 temperature $T_{c0}$ might be the interplane phase slippage promoted
 by thermal motion of pancake vortices inside the layers. However,
 this mechanism does not provide the observed field dependence of the
 resistivity.
A linear field dependence rather suggests a usual flux-flow regime. Of
 course, there is no such thing as flux flow resistivity for current
 flowing along the field direction. Nevertheless, a highly anisotropic
 structure of our $Bi$ samples with alternating quasi-metallic and
 disordered
non-metallic layers favors the current path with the in-plane
 meanders. Then there is a finite  Lorentz force applied to the vortex  
even in  the longitudinal geometry.

It is natural to attribute the high field portion of 
 the curve in Fig.1 (assumed to be above H$_{c2}$) to a normal state 
 magnetoresistance, R$_{N}$(B,T), which appears to be $negative$ and 
 $linear$ in B. The latter is unusual for the longitudinal transport 
 but is also evident in other studies  \cite{and,ong,hus}. 
With these assumptions we can determine (i) the upper critical field, 
H$_{c2}$(T), from the intersection of two linear approximations in 
Fig.1 and (ii) the zero field normal state c-axis resistance, 
R$_{N}$(0,T), by extrapolation of the normal state linear 
magnetoresistance to a zero field.  This procedure 
allows us to separate contributions originating from the normal and 
superconducting states and, in particular, to avoid to large extent an ambiguity due to 
fluctuations in the crossover region.
Referring to Fig.2, the inset shows the field dependence of BSCCO-2212 
 out-of-plane resistance normalised by its normal state field 
 dependence, $R_{N}(B)$, thus accounting for its variation with 
 field and temperature.   The slope of the flux-flow 
 resistance is inversely proportional to $H_{c2}$ as R$_{FF}= 
 R_{N}\times B/H_{c2}$.  Indeed, 
 H$_{c2}$ determined from (i) the intersection of the linear fits 
 mentioned above and (ii) that obtained from $R_{FF}(B)$ as $R_{N}(O,T) 
 (\partial R_{FF}/\partial B)^{-1}$ (Fig.2), are almost identical as is 
 seen from Fig.3 where the temperature dependence of $H_{c2}$ is 
 presented together with the  theoretical fit using the 
 Bose-Einstein condensation critical field \cite{ale} given by
 \begin{equation}
 H_{c2}(T) \sim (t^{-1}-t^{1/2})^{3/2}
 \end{equation}
 with $t=T/T_{c0}$.  
  H$_{c2}$(T) shows an upward 
 temperature dependence in agreement with the previous result based on 
 the low field ($\leq$15 T) scaling of $R(B,T)$ \cite{alezav} and with 
 independent results of other authors \cite{mac,oso,fra,boe,gan}.  We 
 tried (unsuccessfully) to fit the data with the pseudo-upper-critical 
 field, $H^{*}\sim T^{-1}exp(-T/T_{0})$ \cite{ges} as suggested in 
 Ref.  \cite{wen}( the dashed line in Fig.3).  Therefore, 
 the model which lies behind this equation, which is based on 
 Josephson-coupling as the origin of the anomalous H$_{c2}$(T), is not 
 supported by our experiment.  Moreover, there is no change in the 
 temperature dependent slope of resistivity above the superconducting 
 transition as would be the case of superconducting domains formed 
 well above the transition temperature. Some diamagnetism observed
 above the resistive
$T_{c}(B)$\cite{jun2,wen} is explained as the $normal$ state Landau
 diamagnetism of singlet bosons \cite{alekabden}.

The temperature dependence of the zero-field normal state resistance 
$R_{N}(0,T)$ obtained by the linear extrapolation described above, 
together with the theoretical curve (see below) are shown in the main 
panel of Fig.4 by black dots and the solid curve respectively. At low
temperatures 
the high-field range is  too short to define a linear high-field
portion, so that an extrapolation to $B=0$ to obtain directly
$R_N(0,T)$ is not possible. We have found experimentally that a
relation
$R_{N}(0.T)-R_{max}(T) \propto  
\exp(-\alpha T)$ is 
valid  with a constant $\alpha$ in the region where both the maximum value of the resistance
$R_{max}(T)$ (see the  arrows in Fig.1 inset) and $R_{N}(0,T)$ can be 
measured. The 
crosses in Fig.4 are obtained  by  rescaling of $R_{N}(0.T)$ 
according to this relation.  The 
value of a dimensionless negative normal state resistance slope, 
$S(T)= R_{N}(0,T)^{-1}¥\partial R_{N}(B,T) /\partial B$, increases 
rapidly with falling temperature with an upturn around 50 K (the inset 
in Fig.4).

In what follows, we show that the unusual features of the c-axis 
magnetotransport, Figs.  1-4, can be broadly understood within the 
framework of the bipolaron theory \cite{ale,alemot}.  As shown in Fig.3, 
the divergent shape of H$_{c2}(T)$ is 
consistent with the Bose-Einstein condensation field of charged bosons 
pre-formed above T$_{c}$. 
  Within the theory  the  c-axis $normal$-state  transport in 
cuprates is due to single unpaired polarons, except at very low 
temperatures when unpaired carriers are frozen out \cite{alekab}.  Single polarons 
exist as excitations with the energy k$_{B}$T$^{*}=\Delta/2$ or 
larger, where $\Delta$ is the bipolaron binding energy.  The edges of 
two polaronic bands spin-split with respect to the chemical potential 
(pinned at the mobility edge, $\mu\simeq E_{c}\equiv 0$ at low 
temperatures \cite{alekab}) depend on the magnetic field due to a spin 
and orbital magnetic  shifts as
\begin{equation}
{\Delta_{\downarrow,\uparrow}\over{2}}={\Delta\over{2}}+\mu_{B}^{*}B
\pm (J\sigma +\mu_{B}B),
\end{equation}
where $J$ is the exchange interaction of holes with localised copper 
electrons and $\sigma \equiv <S_{i}^{z}/S>$ is an average 
magnetisation of copper per site.  The exchange interaction leads to 
the spin-polarised polaron bands split by $2J\sigma$.  They are 
further split (the last term in Eq.(2)) and shifted by the external 
magnetic field.  Here $\mu_{B}$ and $\mu_{B}^{*}$ are the Bohr 
magnetons determined with the electron $m_{e}$ and polaron $m^{*}$ 
mass, respectively.  We assume that the mobility edge is not affected 
by the magnetic field because bipolarons are heavier than polarons.

Assuming that k$_{B}$T is less than the polaron bandwidth and noting 
that polarons are not degenerate at any temperature,  we obtain for 
the polaron density
\begin{eqnarray}
n_{p}(B,T) &\sim& T^{d/2} \exp[-T^{*}/T-\mu_{B}^{*}B/k_{B}T]\cr
&\times& \cosh
\left[(J\sigma+\mu_{B}B)/k_{B}T\right],
\end{eqnarray}
where d is the dimensionality of the polaron energy spectrum.  The 
exchange splitting of the polaronic bands is responsible for the 
negative magnetoresistance. The characteristic temperature 
$T_{s}\equiv J\sigma/k_{B}$ can be estimated if the magnetisation of 
copper ions is known.  Precise measurements of magnetic neutron 
scattering in high-purity high--T$_{c}$ single crystals of several 
cuprates revealed a rather low number of localised magnetic moments.  
The authors of these experiments noted that "what little scattering is 
observed, corresponds to $\simeq 3.2$ \% of the Cu atoms having a
spin 1/2" \cite{zie}.  Thus, with $\sigma=0.03$ and $J=0.15$ eV we 
obtain $T_{S} \simeq 50$ K. It is reasonable to assume that the 
polaron mobility, $\mu_{p}$, is field and temperature independent in 
the relevant region of B and T because field orbital effects are 
suppressed due to a heavy polaron mass.  The temperature dependence of 
$\mu_{p}$ is almost absent if the scattering is dominated by a random 
potential \cite{alelog}.  As a result one obtains for the temperature 
dependence of the zero field normal state resistivity,
\begin{equation}
R_{N}(T)={R_{0}\over{1+(T/T_{0})^{1/2}\exp (-T^{*}/T)\cosh(T_{s}/T)}},
\end{equation}
where $1/R_{0}$ is a (low) bipolaron c-axis conductivity\cite{alelog}, 
and $T_{0}$ is a parameter depending on the ratio of the bipolaron and 
polaron mobilities.  Here we assume $d=1$ in agreement with the 
angle-resolved photoemission \cite{cam} showing no dispersion along 
certain directions of the two-dimensional Brillouin zone and also with 
the tunnelling spectra successfully described by a one-dimensional DOS 
\cite{aletun}.  The magnetic field slope defined above is then given by
\begin{equation}
S(T)={\mu_{B}[\tanh(T_{s}/T)-m_{e}/m^{*}]\over{k_{B}T}}\left({R_{N}(T)\over{R
_{0}}}-1\right).
\end{equation}
As seen by reference to Fig.4, these expressions are in reasonable 
qualitative agreement with experiment. The solid line in Fig.4 is 
calculated from Eq.(4) using the following values of parameters: 
$R_{0}=1300\,\Omega$, T$^{*}\simeq 170$K , $T_{s}=53\,K$ and 
T$_{0}=0.046\,K$.  The theoretical slope Eq.(5) is calculated 
with the same parameters and with $m_{e}/m^{*}=0.022$. 

Our model of c-axis magnetotransport is supported by other independent
observations. While the  extrapolating procedure might
underestimate the magnitude of the upper critical field, its unusual
temperature dependence is robust as  demonstrated in
the in-plane resistivity data \cite{gan}. The fact that the negative
linear c-axis  magnetoresistance is observed above the zero-field critical
temperature tells us that this unusual phenomenon is a normal state
feature rather than a signature of
some fluctuations in the superconducting state. The measurements of the magnetic
susceptibility \cite{mul} and the doping dependence of superconducting
parameters \cite{hof} support the bipolaron origin of the
normal state pseudogap T$^{*}$.
 The isotope effect
on the normal state  pseudogap observed recently \cite{zur} strongly supports
its bipolaron origin as well. 

 In conclusion, we have  measured the longitudinal out-of-plane 
 magnetoresistance of BSCCO-2212 single crystals in magnetic fields up 
 to 50 T. We found a quasi-linear negative magnetoresistance in the 
 normal state and a quasi-linear positive  magnetoresistance in the 
 mixed state of BSCCO-2212.  This allowed us to determine the upper 
 critical field and to trace the zero field normal state c-axis 
 resistance well below T$_{c0}$.  The shape of H$_{c2}$(T), the 
 temperature dependence of the c-axis resistance and its negative field 
 slope  are understood within the framework of
the 
 bipolaron theory of the cuprates.

This work has been partially supported by EPSRC visiting fellowship
research grant Ref.: GR/L77553, by the Leverhulme Trust, F/182/AT, and
by the
Russian Foundation for Basic Research, GR/98-02-17485.  The authors
greatly appreciate enlightening discussions with Y. Ando, G.
Boebinger, J.R. Cooper, N.E. Hussey,  V.V. Kabanov, W.Y. Liang, and G.
Zhao.  We thank S. Hayden, J.T. Jansenn, P.J. Meeson, R. Murphy, M.
Bennet, and B. Wiltshire for their generous help with
experimental equipment and software.

\centerline{{\bf Figure Captions}}

Fig.1. The  magnetic field dependence of the 
out-of-plane resistance of BSCCO-2212 measured at 78K (T$_{c0}=92$ K).  
Linear fits to the flux-flow portion of the curve and that attributed to 
the normal state magnetoresistance are shown by dashed and solid 
lines respectively.

 The inset shows the variation with field of the normal state 
 magnetoresistance measured at different temperatures, 115, 103, 98, 
 90.1, 78, and 57.5\,K (from the top) normalised by the value of 
 $R_{N}(0)$ (see text).

Fig.2. The  temperature dependence of the flux-flow resistance slope.
Inset: Field dependence of out-of-plane resistance of  BSCCO-2212
normalised by its normal state value, $R_{N}(B)$.  The
selected traces are obtained (from right to the left) at 16, 20, 25, 
30, 35, 45, 52.6, 57.5, 65, 70, 78, and 88.7K respectively.

 Fig.3.  The resistive upper critical field of BSCCO-2212 as a function 
 of temperature obtained from the intersections of the linear 
 extrapolations from the normal and flux-flow regimes (solid circles), 
 and as the ratio of the extrapolated $R_{N}(T)$ and flux-flow
resistance 
 slope (crosses).  A fit to the Bose-Einstein condensation field, 
 Eq.(1), is 
 shown by the solid curve, while the dashed line shows  a fit to the 
 'pseudo-upper-critical field' of Ref. \cite{ges}.

Fig.4. The zero field normal state c-axis 
resistance, $R_{N}(0)$, (main panel)  
and the magnetic field slope, $S(T)$, (inset) of BSCCO-2212 fitted by
Eq.(4) and Eq.(5), 
respectively.  Solid symbols in the main panel correspond to the value 
of $R_{N}(0)$ 
obtained by the linear extrapolation; crosses correspond to the value 
of
$R_{N}(0)$ determined from $R_{max}$, as explained in the text.

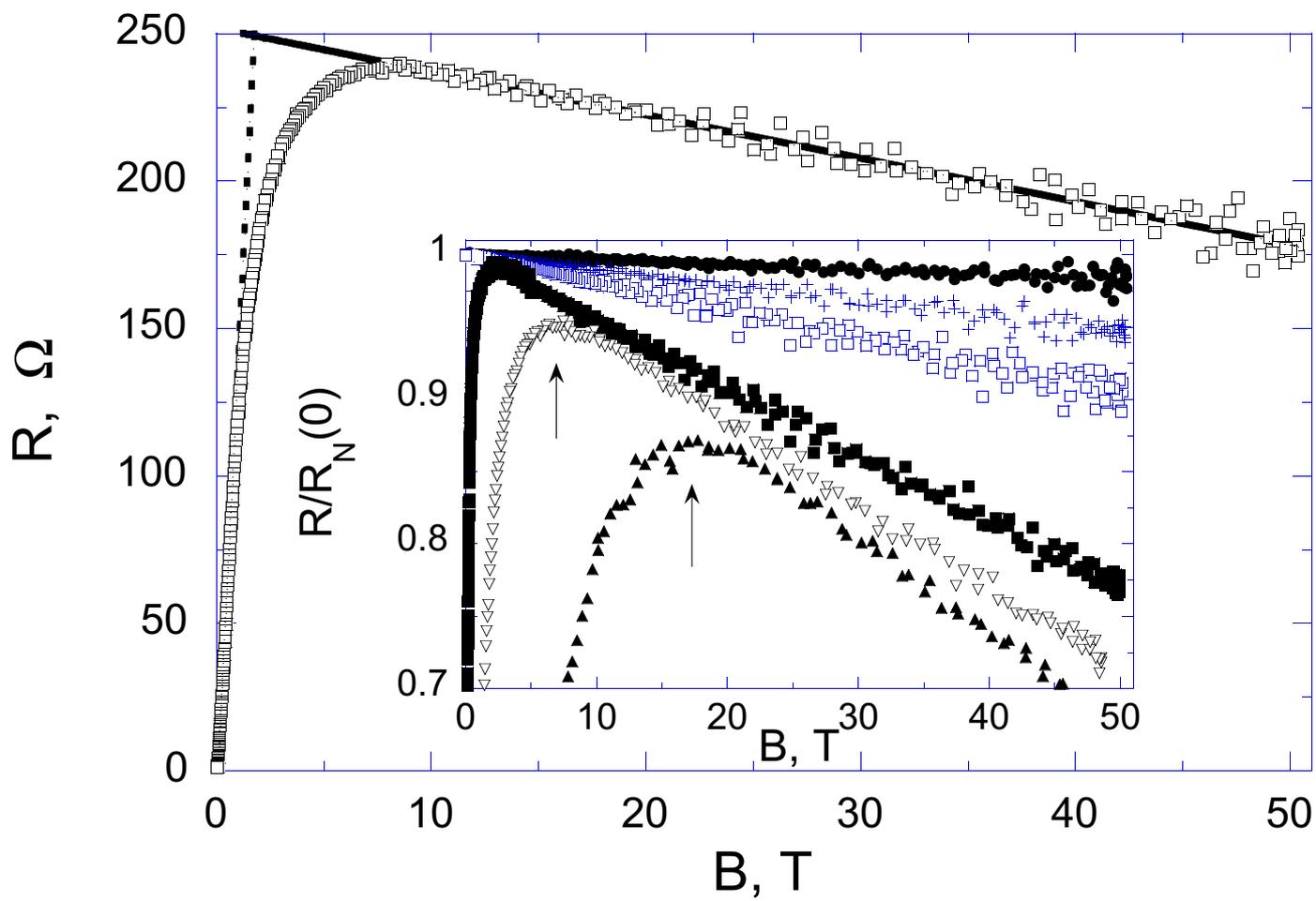
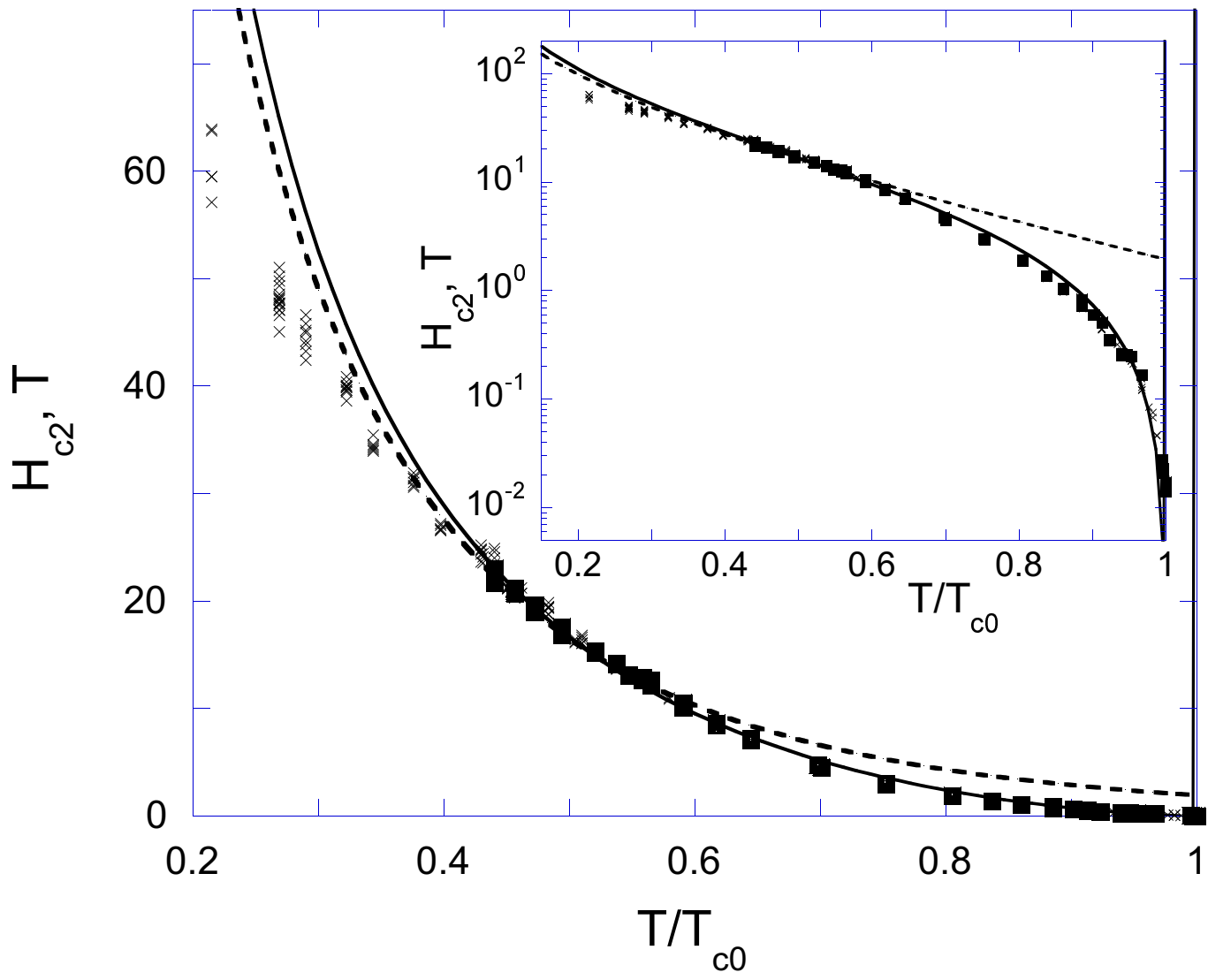
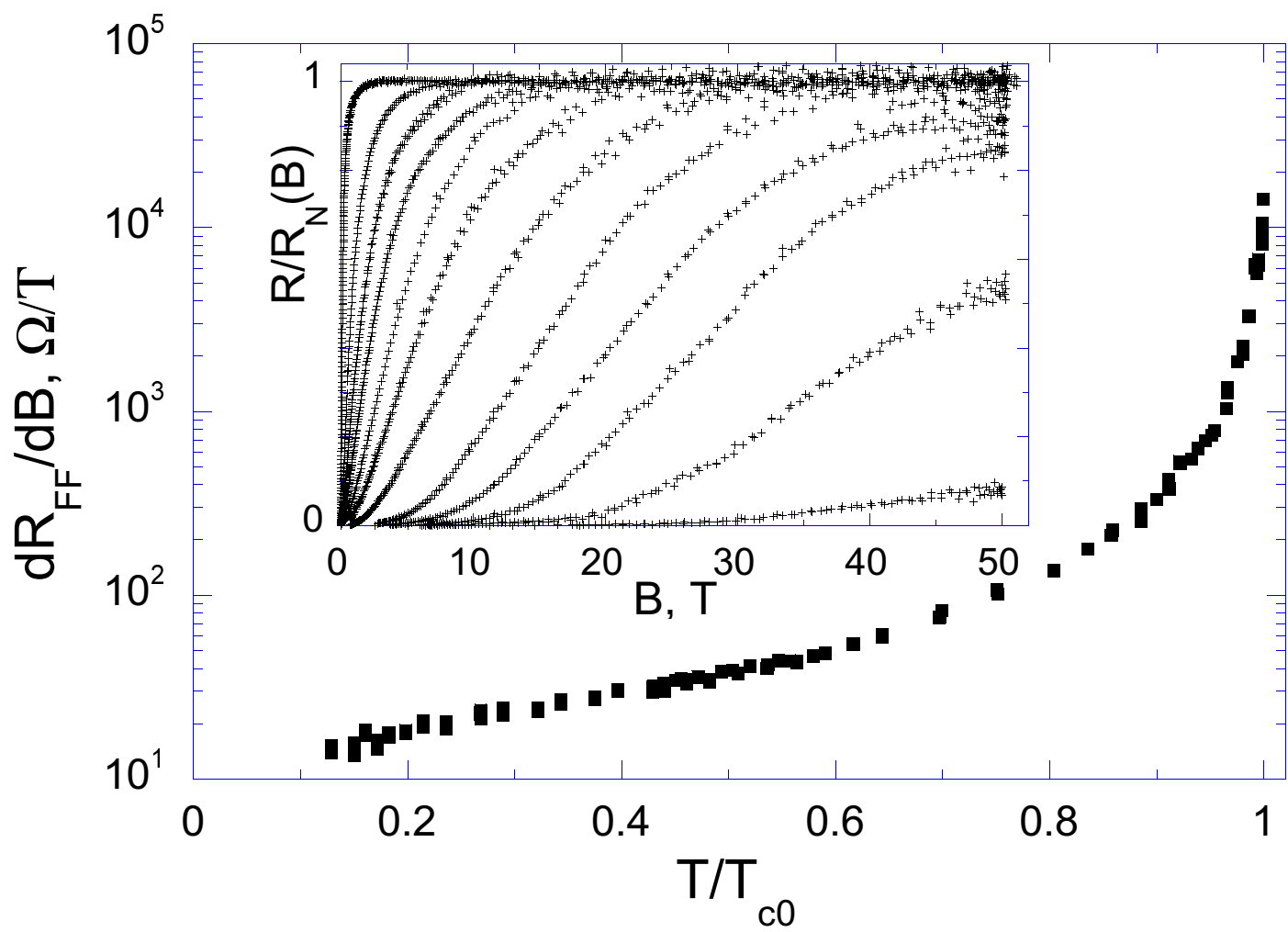
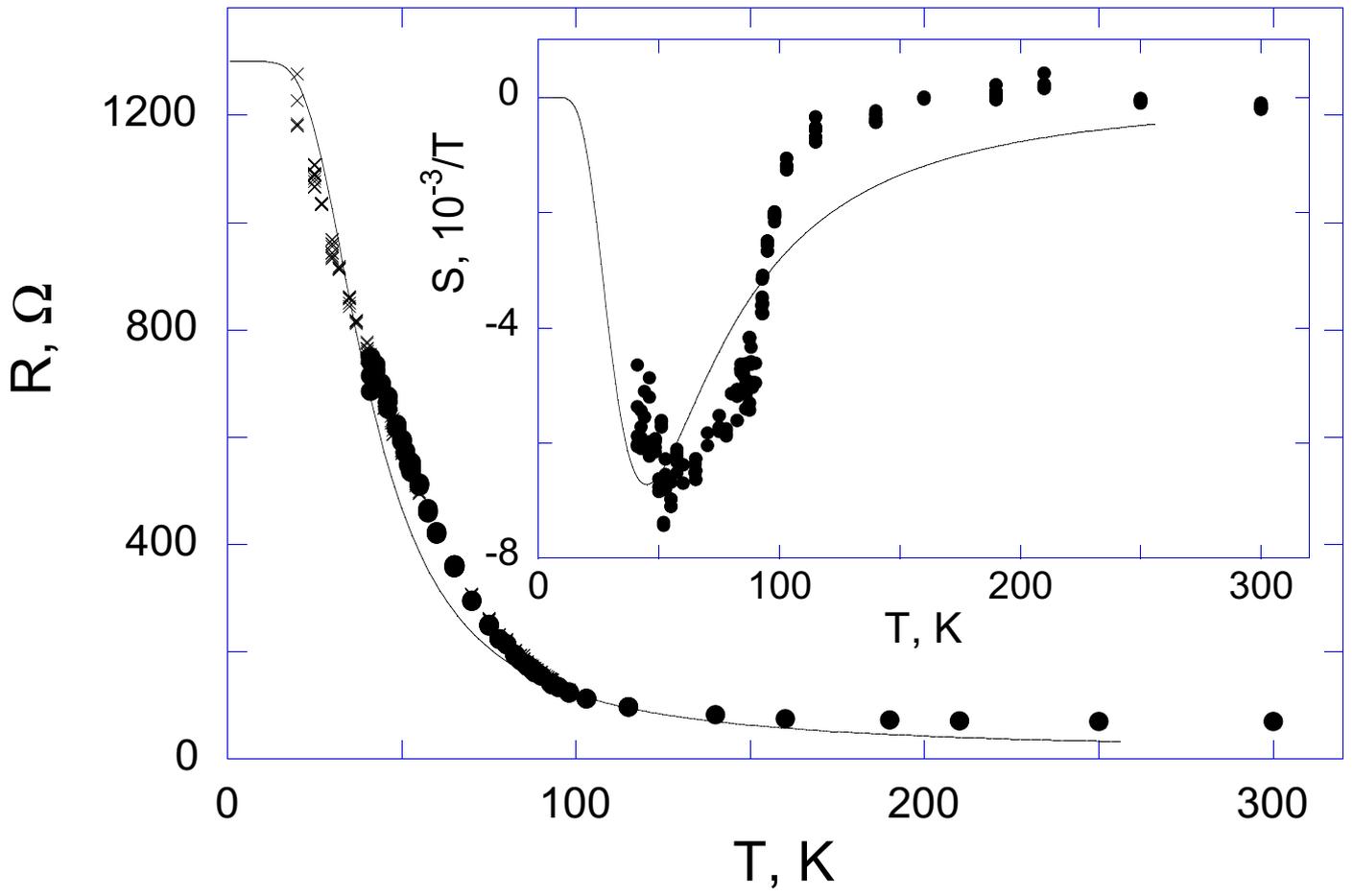
\end{document}